\documentclass[a4paper,fleqn]{cas-dc}

\usepackage[authoryear,longnamesfirst]{natbib}

\def\tsc#1{\csdef{#1}{\textsc{\lowercase{#1}}\xspace}}
\tsc{WGM}
\tsc{QE}

\usepackage{amsmath,amsfonts}
\usepackage{algorithmic}
\usepackage{algorithm}
\usepackage{array}
\usepackage[caption=false,font=normalsize,labelfont=sf,textfont=sf]{subfig}
\usepackage{textcomp}
\usepackage{stfloats}
\usepackage{url}
\usepackage{verbatim}
\usepackage{graphicx}
\usepackage{cite}
\usepackage{xurl}

\usepackage[T1]{fontenc}
\usepackage{graphicx}
\usepackage{amsmath}
\usepackage[caption=false]{subfig}
\usepackage{caption}
\usepackage{tcolorbox}
\tcbuselibrary{breakable, skins}
\usepackage{makecell}
\usepackage[table]{xcolor} 
\usepackage{pifont}
\usepackage{xcolor}
\usepackage[table]{xcolor}
\usepackage{tabularx}
\usepackage{booktabs}

\newcommand{\cmark}{\textcolor{green!60!black}{\ding{51}}} 
\newcommand{\xmark}{\textcolor{red}{\ding{55}}}            
\newcommand{\pmark}{\textcolor{orange}{\ding{115}}}        

\begin{document}
\let\WriteBookmarks\relax
\def\floatpagepagefraction{1}
\def\textpagefraction{.001}

\shorttitle{}    

\shortauthors{}  

\title [mode = title]{MalGEN: A Testbed for Modeling and Evaluating Malware Behaviors}



%

\author[1]{Bikash Saha}

\cormark[1]

\fnmark[1]

\ead{bikash@cse.iitk.ac.in }

\ead[url]{}

\credit{Conceptualization of this study, Data Curation, Investigation, Methodology, Writing - Original Draft, Project administration, Writing - Review \& Editing, Visualization}

\affiliation[1]{organization={Indian Institute of Technology Kanpur},
            country={India}}

\author[2]{Sandeep Kumar Shukla}

\fnmark[2]

\ead{sandeeps@iiit.ac.in}

\ead[url]{}

\credit{Resources, Validation, Writing - Review \& Editing, Supervision}

\affiliation[2]{organization={International Institute of Information Technology, Hyderabad},
            country={India}}

\cortext[1]{Corresponding author}

\fntext[1]{}


\begin{abstract}
Modern cybersecurity requires systematic ways to evaluate how detection systems respond to evolving and previously unseen attack behaviors. Existing malware repositories largely capture known patterns and provide limited support for stress-testing defenses against novel threats. To address this, we present MalGEN, a modular testbed that models adversarial workflows and generates executable artifacts in a controlled environment. The framework decomposes high-level attack objectives into structured stages, enabling the synthesis of diverse and multi-stage behaviors. We evaluate MalGEN across 1,920 benchmark settings covering multiple platforms and behavioral objectives, resulting in 977 executable samples. Analysis shows that the generated artifacts exhibit a wide range of malicious techniques and multi-stage attack patterns. However, 45.71\% of these samples remain undetected by existing detection engines, which reveals notable gaps in current defenses. These findings provide practical insights into the limitations of widely used detection approaches and support the development of more robust security evaluation and testing practices.
\end{abstract}




\maketitle

\section{Introduction}
\label{sec:MalGEN_intro}

The escalating complexity of modern cyber threats has placed defenders under increasing pressure to counter rapidly evolving adversarial behaviors~\citep{enisa2025threatlandscape}. Despite continuous advances in security solutions, adversaries persistently craft sophisticated malware that adapts to and evades existing defenses~\citep{geng2024survey,afianian2019malware,demetrio2021adversarial}. This ongoing offense–defense race highlights the need to move beyond purely reactive strategies toward proactive approaches that can anticipate emerging threats. Traditional detection systems, while effective against known patterns, remain fundamentally limited in capturing previously unseen behaviors~\citep{hidayat2025advancements,gorment2023machine}. 


Importantly, this challenge does not stem from a lack of malware data. Large-scale repositories provide extensive collections of real-world samples, primarily capturing known malicious behaviors. As a result, detection systems trained and evaluated on such data are inherently biased toward known threat patterns and often struggle to generalize to novel or emerging behaviors~\citep{jordaney2017transcend,barbero2022transcending}. This highlights the need for frameworks that enable controlled modeling of novel adversarial behaviors for effective stress-testing of detection systems.


Recent advances in program synthesis and automated code generation provide new opportunities to model adversarial behaviors in controlled settings~\citep{naveed2025comprehensive,achiam2023gpt}. These capabilities allow the construction of realistic attack workflows that can be used to study emerging threat patterns and evaluate defensive systems. These capabilities are already being applied to cybersecurity tasks such as threat intelligence extraction, vulnerability analysis, and incident response~\citep{xu2024large,sheng2025llms,saha2025parag,saha2025malaware,tellache2025advancing}. Beyond their defensive utility, LLMs can also be leveraged in controlled settings to model adversarial behaviors, enabling systematic exploration of emerging attack patterns. Similar to red teaming, such controlled simulation provides a practical means to study attacker strategies and evaluate defensive readiness.

However, existing efforts on LLM-driven malware generation remain limited in both scale and structure~\citep{adamec2024development,yamin2024combining,li2025automated,madani2023metamorphic}. Most approaches focus on single-step or template-based code generation and fail to capture the multi-stage reasoning, coordination, and adaptability observed in real-world attacks. Moreover, current benchmarks and evaluation practices are largely based on static datasets, which do not adequately reflect AI-generated, behaviorally diverse threats. This creates a gap between rapidly evolving generative capabilities and existing evaluation methodologies.

To address these limitations, we present MalGEN, a modular testbed for generating and evaluating behaviorally diverse malware in a controlled environment. MalGEN models adversarial workflows as a sequence of stages, including intent decomposition, payload construction, and integration, enabling the creation of multi-stage attack behaviors. The process is guided by real-world adversarial patterns informed by frameworks such as MITRE ATT\&CK~\citep{strom2018mitre}. By varying task structures and execution paths, MalGEN produces structurally diverse binaries that are not well captured by existing detection systems.




We evaluate MalGEN across eight code-oriented LLM variants, two operating systems, and 120 adversarial objectives, resulting in 1,920 configurations. The framework generates executable samples with multi-stage behaviors and varying complexity. We analyze these artifacts in terms of behavioral characteristics, execution outcomes, and detection patterns, highlighting gaps in current detection approaches.
In summary, we make the following contributions:

\begin{itemize}
    \item We introduce an intent-driven malware generation process that converts high-level adversarial goals into smaller tasks and combines them to create diverse malware behaviors.

    \item We present \textsc{MalGEN}, a modular testbed that uses multiple agents to plan tasks, generate code, integrate the generated code, and build executable malware samples.

    \item We evaluate \textsc{MalGEN} across 1,920 benchmark settings using eight code-oriented LLM variants, 120 adversarial behavior prompts, and two operating systems, Linux and Windows.

    \item We analyze 977 generated executable samples using MITRE ATT\&CK techniques and tactics to study their behavioral diversity and multi-stage attack patterns.

    \item We evaluate the generated samples using VirusTotal detection engines and behavior-based machine learning detectors to understand how existing detection systems respond to AI-generated malware.

    \item We analyze the capabilities and failure cases of the generated samples to understand what types of malware behaviors are produced and where the generation process still fails.
\end{itemize}


\noindent The remainder of this paper is structured as follows. Section~\ref{sec:MalGEN_Motivation} outlines the motivation and objectives of this research. Section~\ref{sec:MalGEN} describes the architecture and implementation details of the proposed framework. Section~\ref{sec:experiments} explains the experimental methodology adopted for extensive evaluation. Section~\ref{sec:results} presents the key results, discussions, and insights derived from the experiments. Section~\ref{sec:MalGEN_limitations_future_work} highlights the limitations of this study and outlines potential directions for future work. Section~\ref{sec:relatedwork} reviews related research on AI-assisted adversarial behavior generation. Finally, Section~\ref{sec:conclusion} concludes the paper.

\section{Ethics Considerations}
\label{sec:ethics}
MalGEN is developed and intended strictly for academic, defensive, and educational research. Its use must occur within controlled research environments, such as virtualized sandboxes or non-networked testbeds, to prevent unintended propagation or system impact. Users are expected to adhere to ethical AI and responsible disclosure principles, ensuring that all experimentation remains transparent, contained, and compliant with institutional review standards.

The framework’s purpose is to advance understanding of adversarial behaviors and strengthen defensive readiness, not to deploy operational malware. Any use of MalGEN for offensive, destructive, or commercial purposes is explicitly prohibited. These ethical commitments align with the risk-mitigation and responsible innovation principles outlined in \citep{brundage2018malicious,shevlane2023model}, ensuring that MalGEN contributes constructively to the security research ecosystem while minimizing potential misuse.

\section{Motivation}
\label{sec:MalGEN_Motivation}
\begin{table*}[b]
    \centering
    \caption{Motivation--gap--solution overview}
    \label{tab:malgen_motivation_gap}
    \begin{tabularx}{\textwidth}{lXX}
        \hline
         \textbf{Existing State} & \textbf{Research Gap} & \textbf{MalGEN Approach} \\
         \hline
         Single-step code synthesis demonstration & Absence of end-to-end adversarial workflow modeling & Multi-agent orchestration of full attack chain: \emph{(Adversarial Intent }$\rightarrow$ \emph{Develop }$\rightarrow$ \emph{Payload }$\rightarrow$ \emph{Executable)} \\
         Legacy dataset; static payload & Lack of benchmarks capturing novel, AI-synthesized behaviors & Generates diverse, functionally rich samples; supports red-teaming and resilience testing \\
         Weak behavior realism & Detection systems trained on outdated behaviors & Produces behaviorally faithful, ATT\&CK-aligned artifacts for realistic evaluation \\ \bottomrule
    \end{tabularx}
\end{table*}
The growing sophistication of cyber threats has exposed fundamental limitations in traditional defense paradigms. Despite advances in machine learning–based detection systems, most approaches still rely on historical datasets, static features, or rule-based signatures that fail to capture evolving adversarial behaviors~\citep{tayyab2022survey,aboaoja2022malware,hidayat2025advancements}. As a result, defenders often face blind spots when confronted with novel malware behavior. These limitations underscore the urgent need for research frameworks that can model and simulate emerging attack behaviors before they manifest in the wild.

Generative artificial intelligence, particularly LLMs, offers an unprecedented opportunity to address this challenge by modeling adversarial intent and reasoning patterns in ways that traditional data-driven approaches cannot~\citep{yamin2024combining, beckerich2023ratgpt, pa2023attacker}. By generating synthetic yet behaviorally rich malware samples, generative models can help researchers explore attack pathways, test defensive resilience, and identify unseen detection blind spots in a controlled environment. However, without a structured testbed to orchestrate such modeling, these benefits remain largely untapped.

Existing studies exploring LLM-driven malware remain limited in both scale and depth. Most focus narrowly on single-step code generation, overlooking the multi-stage reasoning, coordination, and adaptation that characterize real-world intrusion campaigns~\citep{adamec2024development,yamin2024combining,li2025automated, madani2023metamorphic}. Equally, today’s malware detection benchmarks are built on legacy datasets that lack representation of AI-synthesized, evasive and novel behaviors. This mismatch between rapidly evolving AI-enabled threats and static evaluation practices leaves defenders underprepared for the emerging generation of adversarial tactics and techniques.


For security researchers, red teams, and developers of detection engines, the central challenge is not simply generating malicious code but understanding the attacker’s operational logic. Modern adversaries do not act in isolation; they plan objectives, select tactics, develop payloads, and evade defenses iteratively~\citep{strom2018mitre}. Capturing this multi-stage reasoning process is essential for developing defenses that can anticipate, rather than merely react to evolving threats. 
Table~\ref{tab:malgen_motivation_gap} summarizes the existing state, corresponding research gaps, and how MalGEN addresses them.




MalGEN addresses these intertwined gaps by providing a purpose-built, agentic testbed for the controlled simulation of LLM-generated malware behaviors. Unlike prior tools that stop at code synthesis, MalGEN captures the full adversarial workflow, from intent formulation to capability planning, payload creation, and evasion strategy design. By modeling attacker reasoning within a multi-agent architecture, MalGEN models behaviorally rich malware samples for testing and strengthening defense systems.

Ultimately, the motivation behind MalGEN lies in transforming a potential risk into a research asset: leveraging the offensive capabilities of LLMs to empower defensive innovation. By enabling controlled red-teaming, behavior-aware benchmark generation, and systematic evaluation against AI-enabled threats, MalGEN bridges the current gap between theoretical awareness and practical preparedness in the era of generative cybersecurity.

\section{MalGEN}
\label{sec:MalGEN}

\begin{figure*}[htbp]
    \centering
    \includegraphics[width=\textwidth]{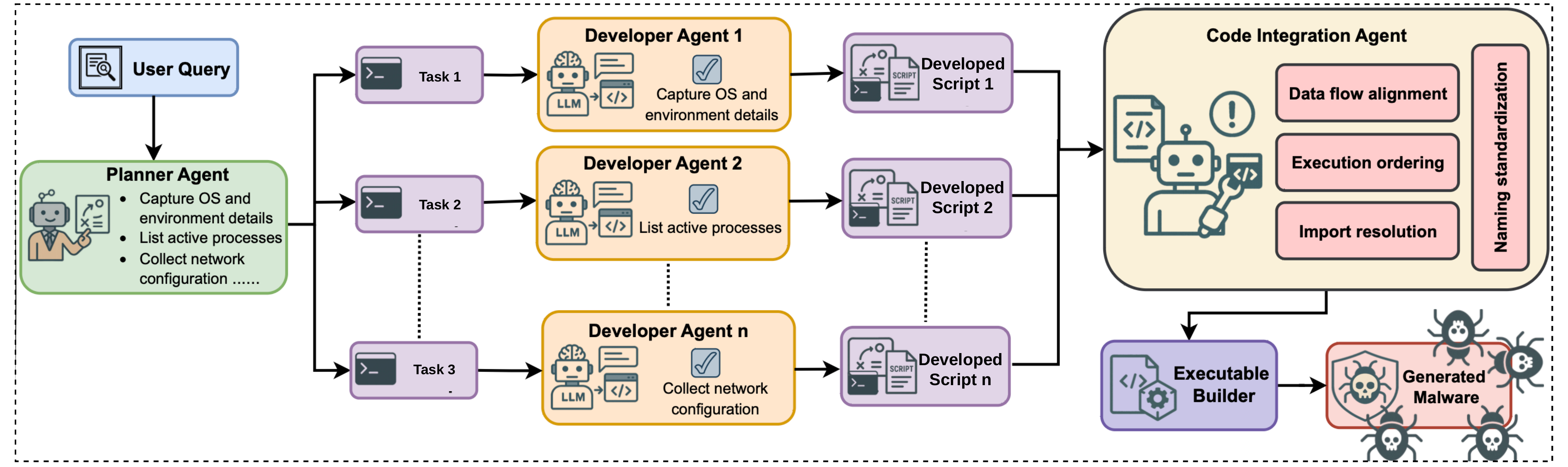}
    \caption{MalGEN architecture showing modular agent interactions across the malware generation pipeline.}
    \label{fig:malgen_architecture}
\end{figure*}

MalGEN is a generative AI–driven testbed for modeling LLM-generated malware behaviors in a controlled and ethically bounded environment. It transforms high-level adversarial intents into executable artifacts through a structured, multi-stage process, where adversarial intent are decomposed, modeled, and integrated into behaviorally diverse malicious binaries. This modular, multi-agent design enables realistic modeling of adversarial workflows while maintaining scalability and transparency. The overall architecture and data flow of MalGEN are shown in Figure~\ref{fig:malgen_architecture}.

\subsection{Agent-Based Architecture}
\label{sec:malgen_agent_arch}

MalGEN follows a staged, agent-based design in which multiple LLM-powered components cooperate to construct malware from high-level adversarial intent. The system progressively transforms this intent into an executable artifact through planning, code generation, integration, and packaging stages. This modular separation improves interpretability and debuggability, while enabling individual components to be independently modified or extended.

\subsubsection{Task Planner Agent}
\label{sec:task_planner}

The Task Planner acts as the entry point of the pipeline. Given a high-level adversarial intent, such as ``collecting system information and exfiltrating it", the planner converts it into a structured execution plan. It defines what needs to be done and in what order. Each sub-task represents a specific functionality and includes its dependencies on other steps.

The planner is implemented using an LLM-based agent with constrained prompting to produce consistent, machine-readable outputs. This ensures that the decomposition remains interpretable and reproducible. Importantly, the same high-level objective can be decomposed in multiple valid ways, leading to different execution paths. This variation contributes to behavioral diversity by producing different combinations and execution paths for the same high-level objective, resulting in behaviors that do not align with existing detection patterns.

\begin{center}
\begin{tcolorbox}[
  breakable,
  enhanced,
  colback=gray!5,
  colframe=black,
  title={Task Planner Agent}
  ]
Decomposes high-level adversarial intent into an ordered sequence of dependent sub-tasks to enable controlled and interpretable behavior generation.
\end{tcolorbox}
\end{center}

\subsubsection{Developer Agent}
\label{sec:developer_agent}

The Developer Agent translates each sub-task into executable code. It takes the structured plan and generates one code snippet per sub-task using an LLM-based code generation model. The prompts clearly specify the intended functionality, execution environment, and required libraries, ensuring that the generated code aligns with the planned behavior.
Each snippet is designed to be modular and self-contained, which simplifies both integration and debugging. This also allows failures to be traced back to individual sub-tasks instead of the entire program.

\begin{center}
\begin{tcolorbox}[
  breakable,
  enhanced,
  colback=gray!5,
  colframe=black,
  title={Developer Agent}
  ]
Generates modular code snippets for each planned sub-task to ensure functional correctness and compatibility with the target environment.
\end{tcolorbox}
\end{center}

\subsubsection{Code Integration Agent}
\label{sec:integration_agent}

The Code Integration Agent combines the independently generated snippets into a single executable script. Since the code is generated in parts, inconsistencies may arise. The integration step resolves such issues by aligning variable and function names, removing redundant imports, and ensuring that execution follows the dependencies defined in the plan.
In addition, the agent preserves data flow between sub-tasks so that outputs from earlier stages are correctly used in later ones. Basic validation checks, such as syntax verification, are applied to detect integration errors. 

\begin{center}
\begin{tcolorbox}[
  breakable,
  enhanced,
  colback=gray!5,
  colframe=black,
  title={Code Integration Agent}
  ]
Combines modular snippets into a consistent and executable script by resolving dependencies and preserving execution flow.
\end{tcolorbox}
\end{center}


\subsubsection{Executable Builder}
\label{sec:executable_builder}

The Executable Builder converts the integrated script into a standalone binary using standard packaging tools. This ensures that the generated artifact can run in a controlled environment without external dependencies.
The resulting binaries can be directly used in sandboxed testbeds or analysis pipelines. This enables realistic evaluation of their behavior and interaction with detection systems.

\begin{center}
\begin{tcolorbox}[
  breakable,
  enhanced,
  colback=gray!5,
  colframe=black,
  title={Executable Builder}
  ]
Packages the integrated script into a standalone executable, enabling controlled execution and evaluation.
\end{tcolorbox}
\end{center}

\noindent This structured pipeline ensures interpretability at every stage, supports modular replacement of components, maintains comprehensive logging for traceability and debugging, and facilitates reproducible, ethically constrained experimentation in malware behavior modeling.

\section{Experiments}
\label{sec:experiments}

This section describes the experimental setup and evaluation criteria used to assess MalGEN. The experiments focus on three main objectives:  
(1) evaluating whether MalGEN can generate functional, executable malware-like artifacts from high-level adversarial intents;  
(2) examining the behavioral realism, diversity, and malicious capabilities of the generated samples; and  
(3) assessing how these artifacts are detected by commercial antivirus and behavioral analysis systems.


\subsection{Implementation Details}
\label{sec:design_choices}


MalGEN is implemented to support controlled generation and reproducible evaluation of adversarial behaviors. We use compact, instruction-tuned language models for code generation and evaluate the framework across eight representative open-source code-oriented LLMs\footnote{Closed-source models were not considered, as they often restrict responses involving potentially malicious instructions.} from different families, including
\texttt{Qwen2.5-Coder-3B/7B}, \texttt{CodeLlama-7B}, \texttt{DeepSeek-Coder-1.3B/6.7B/7B-v1.5}, \texttt{StarCoder-3B}, and \texttt{CodeGemma-7B}. All models are executed locally with identical inference settings (\texttt{temperature}=0, \texttt{top\_p}=1) to enable deterministic decoding and consistent outputs across runs. Python is used as the target language due to its simplicity, auditability, and compatibility with controlled experimentation environments.

MalGEN adopts a modular and context-aware design in which components operate independently across stages such as planning, code generation, integration, and packaging, enabling easier debugging, flexible replacement, and isolated evaluation. Code generation is guided by prompts incorporating task objectives, execution constraints, and relevant libraries, ensuring alignment with intended behaviors while improving consistency and traceability.

\subsection{Dataset and Benchmark Generation}
\label{subsec:dataset_generation}

We construct a set of $120$ high-level adversarial behavior descriptions, each capturing multi-stage objectives that reflect common attacker workflows (e.g., reconnaissance $\rightarrow$ persistence $\rightarrow$ exfiltration). The templates are written in natural language and intentionally abstracted to avoid exploit-specific instructions.

To evaluate variability across models and platforms, we generate artifacts for two target operating systems (Linux and Windows) using eight code-oriented LLMs. For each behavior description, every model produces one artifact per OS, resulting in a total of $120 \times 8 \times 2 = 1{,}920$ benchmark instances.

The resulting corpus is organized by operating system and model family. Each artifact is accompanied by metadata to ensure traceability, including model identifier, prompt templates, generation parameters, source code logs, build configuration, packaging details, build hashes, and SHA-256 digests of the final binaries. All artifacts are generated and executed exclusively within isolated sandbox environments.

\subsection{Experimental Design and Evaluation Environment}

We evaluate MalGEN through controlled experiments that examine how high-level adversarial intents are translated into executable artifacts. For each objective, the framework produces end-to-end samples exhibiting multi-stage behaviors. Each experiment yields a self-contained artifact that is subsequently analyzed to assess its behavioral characteristics, malicious capabilities, and interaction with existing detection systems.

All experiments were conducted in fully isolated, non-networked sandbox environments to ensure safe execution and containment of generated artifacts. Sandboxes were reset after each run to avoid state carryover. Each executable was analyzed using both static and dynamic approaches. For static analysis, samples were submitted to VirusTotal (VT), which aggregates results from multiple antivirus engines and primarily reflects signature-based detection. To complement this, dynamic analysis was performed using Cuckoo Sandbox, where samples were executed to capture system-level behaviors. Results from both analyses were aggregated to assess detection coverage and behavioral variation. 
In addition to behavioral inspection, the Cuckoo-generated API traces were also used for a downstream dynamic malware detection experiment. Specifically, we evaluated whether standard machine learning classifiers trained on a public API-call benchmark could recognize MalGEN-generated samples as malicious. This additional setup enables us to assess not only the behaviors exhibited by generated binaries, but also how well existing behavior-based detectors generalize to them.

\subsection{Analytical Methods}
\label{subsec:analytical_methods}

\begin{figure*}[htbp]
    \centering
    \includegraphics[width=11cm]{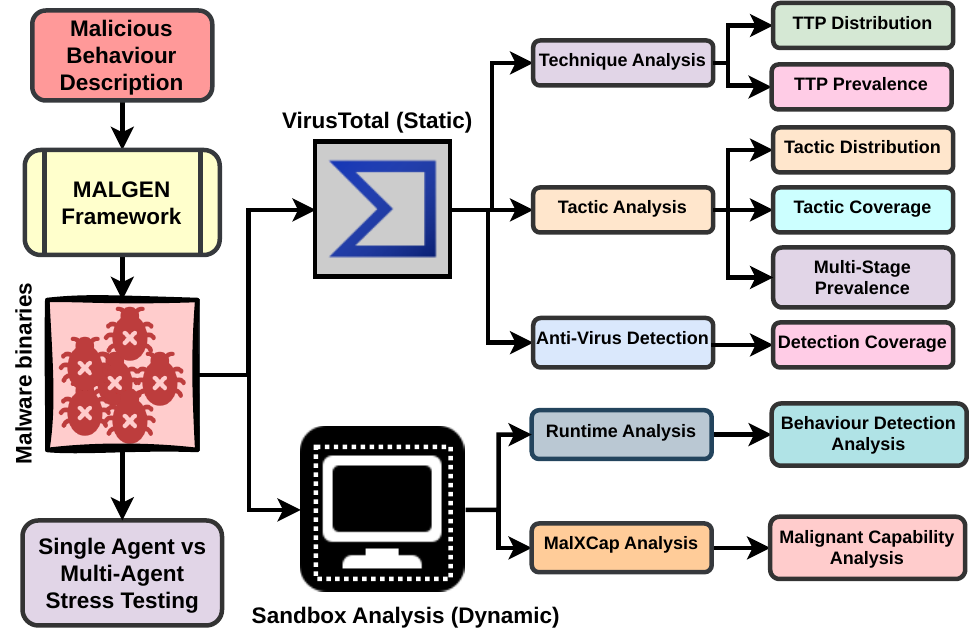}
    \caption{Overall Experiment Performed.}
    \label{fig:malgen_experiemtns}
\end{figure*}

We analyze the generated benchmark using a set of complementary measurements that capture generation success, behavioral characteristics, and detection outcomes. Figure~\ref{fig:malgen_experiemtns} demonstrate the flow of experiments performed for this extensive analysis.
All metrics are computed at the sample level and then aggregated by LLM variant and behavior category.

\paragraph{Sample generation analysis.}
We measure the ability of MalGEN to produce executable artifacts by computing the binary generation success rate across all benchmark settings. For each configuration, we record whether the pipeline successfully produces a functional executable after code generation, integration, and packaging. To assess the role of the multi-agent design, we compare this setup with a single-agent baseline, where a single LLM directly generates code from the input description.

\paragraph{Technique-level analysis.}
To characterize the behaviors present in the generated samples, we extract ATT\&CK technique labels from analysis reports. We analyze both the distribution of distinct techniques per sample to capture behavioral complexity and the overall prevalence of techniques across the corpus to identify commonly represented behaviors. These analyses enable comparison of behavioral diversity and concentration across models.


\paragraph{Tactic-level analysis.}
We map extracted techniques to their corresponding ATT\&CK tactics and analyze both the number of distinct tactics per sample and the overall proportion of tactics represented across generated samples. This enables assessment of how effectively different models cover multiple stages of the attack lifecycle.


\paragraph{Multi-stage prevalence analysis.}
We measure the extent to which samples exhibit multi-stage behavior by counting the number of distinct tactics per sample and computing the proportion of samples spanning multiple stages.

\paragraph{Detection coverage analysis.}
We evaluate how generated samples are flagged by antivirus engines using VirusTotal. For each sample, we compute the detection rate as the ratio of detection count to the number of engines reporting on that sample.

From these values, we derive summary statistics per model, including the fraction of undetected samples, mean detection rate, median detection count, and maximum detection count. These metrics are used to compare how generated samples are handled by existing detection systems.

\paragraph{Dynamic Behavior-Based Detection Analysis}
\label{subsec:dynamic_detection}
To complement VirusTotal-based detection coverage, we evaluate whether standard machine learning based dynamic malware detectors trained on public sandbox~\citep{pikker_sandbox} system call traces can recognize the generated samples as malicious. 
We construct a benchmark corpus from malware  containing 1,000 benign and 1,000 malware samples for both Windows and Linux samples~\citep{tqqm-aq14-19,ygtk-hv95-25}. Further, we use stratified 80:20 train-test split. Each sample is represented using TF-IDF features over API-call  bigrams and trigram, with a maximum vocabulary size of 50,000 features. We evaluate seven standard classifiers: Logistic Regression, Random Forest, K-Nearest Neighbors, Linear SVM, Multinomial Naive Bayes, Decision Trees, and XGBoost. After benchmarking on the held-out split, the trained models are applied directly, without retraining, to API call traces from the generated samples. This step evaluates detection performance using feature sets that are commonly used in machine learning based malware detection systems.


\paragraph{Capability analysis.}
To examine malignant functional behavior, we perform capability-level analysis using the MalXCap extractor~\citep{saha2023malxcap}. Samples are executed in a sandbox environment, and runtime traces are used to extract observed capabilities. These are aggregated to estimate how frequently different behaviors appear across the generated corpus.

\section{Results and Discussion}
\label{sec:results}
Building on the comprehensive experimental plan outlined in Section~\ref{sec:experiments}, we conducted an extensive evaluation of the MalGEN framework. This section presents the results obtained from each analysis and discusses the key observations and insights derived from the corresponding experiments.

\subsection{Sample Generation Analysis}

We executed MalGEN across $1{,}920$ benchmark settings, as described in Section~\ref{subsec:dataset_generation}. The framework successfully generated $977$ executable artifacts, corresponding to an overall success rate of $50.89\%$. The generated samples are approximately balanced across Linux and Windows platforms. Table~\ref{tab:single_vs_multi} compares binary generation performance in a single-agent setting with the proposed multi-agent pipeline.

The single-agent setting shows clear limitations in producing executable binaries. When a single model is asked to handle the full task directly, failures arise from incomplete logic, missing imports, unresolved dependencies, or inconsistent code structure. This is particularly visible for models such as CodeLlama-7B and Qwen2.5-Coder-3B, where executable generation remains relatively low.

\begin{table*}[htbp]
\centering
\caption{Single-Agent vs. Multi-Agent Binary Generation Benchmark}
\label{tab:single_vs_multi}
\small
\begin{tabular}{|l|c|c|c|c|}
\hline
\textbf{LLM Variant} & \multicolumn{2}{c|}{\textbf{Single-Agent}} & \multicolumn{2}{c|}{\textbf{Multi-Agent}} \\
\cline{2-5}
 & \textbf{Exec (\%)} & \textbf{Fail (\%)} & \textbf{Exec (\%)} & \textbf{Fail (\%)} \\
\midrule
\hline
CodeGemma-7b & 28.33 & 71.67 & 50.83 & 49.17 \\

CodeLlama-7b & 20.83 & 79.17 & 35.42 & 64.58 \\

Deepseek-Coder-1.3b & 31.67 & 68.33 & 52.08 & 47.92 \\

Deepseek-Coder-6.7b  & 33.33 & 66.67 & 52.92 & 47.08 \\

Deepseek-Coder-7b-v1.5 & 27.50 & 72.50 & 45.83 & 54.17 \\

StarCoder-3b & 25.00 & 75.00 & 49.17 & 50.83 \\

Qwen2.5-Coder-3b  & 23.33 & 76.67 & 39.58 & 60.42 \\

\textbf{Qwen2.5-Coder-7b} & \textbf{52.50} & \textbf{47.50} & \textbf{81.25} & \textbf{18.75} \\
\midrule
\hline
\textbf{Overall (Average)} & \textbf{30.31} & \textbf{69.69} & \textbf{50.89} & \textbf{49.11} \\
\hline
\end{tabular}
\end{table*}

The multi-agent pipeline improves this process by breaking generation into intermediate stages, including planning, modular development, and integration. This reduces the burden on any single model and makes the overall workflow more stable. As shown in Table~\ref{tab:single_vs_multi}, executable generation improves for every model under the multi-agent setting.

On average across models, executable success increases from $30.31\%$ in the single-agent setup to $50.89\%$ in the multi-agent setup, while the corresponding failure rate drops from $69.69\%$ to $49.11\%$. The gains are consistent across model families. For example, CodeGemma-7B improves from $28.33\%$ to $50.83\%$, and DeepSeek-Coder-1.3B improves from $31.67\%$ to $52.08\%$. The strongest result is obtained by Qwen2.5-Coder-7B, which reaches $81.25\%$ executability under the multi-agent pipeline, compared to $52.50\%$ in the single-agent setting.

Overall, these results indicate that structured multi-stage generation substantially improves binary generation reliability. This improvement is observed not only for stronger models, but also for smaller and mid-sized ones, suggesting that the pipeline design plays an important role in successful end-to-end generation.

\begin{center}
\begin{tcolorbox}[
  breakable,
  enhanced,
  colback=gray!5,
  colframe=black,
  title={Key Observation \#1: Sample Generation Analysis}
  ]
The multi-agent pipeline improves executable generation across all evaluated models by introducing structured intermediate stages, leading to higher success rates and fewer failures than the single-agent setup.
\end{tcolorbox}
\end{center}

\subsection{Technique-Level Analysis}
\label{subsec:technique_analysis_results}
To characterize the behavioral patterns in generated samples, we analyze both the distribution of distinct MITRE ATT\&CK techniques per sample and the prevalence of techniques across the corpus. 

\begin{figure}
    \centering
    \includegraphics[width=\columnwidth,keepaspectratio]{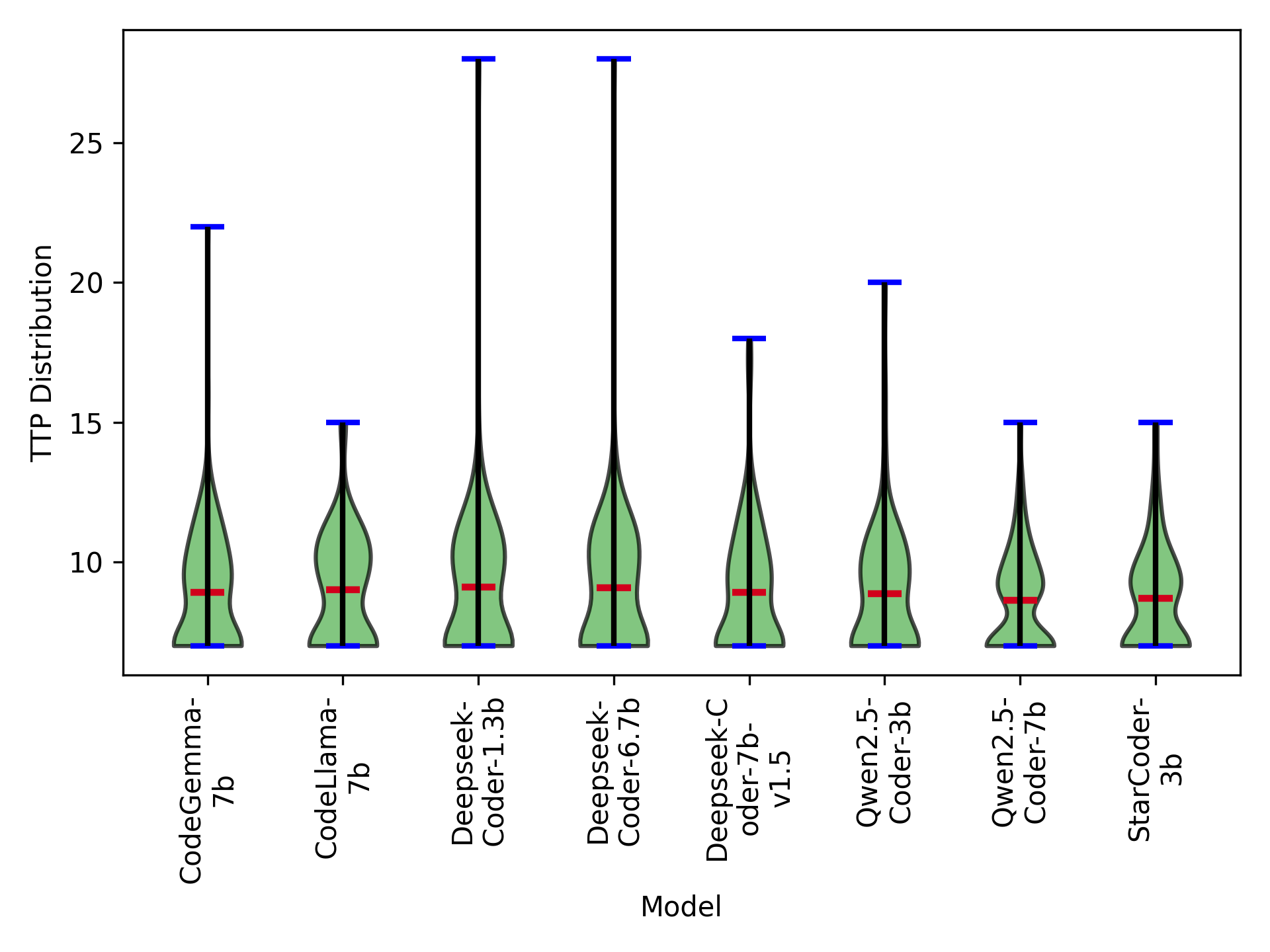}
    \caption{Distribution of techniques per sample.}
    \label{fig:ttp_violin}
\end{figure}

Figure~\ref{fig:ttp_violin} shows the distribution of distinct techniques observed per generated sample across LLM variants. Each violin represents the density of TTP counts per sample, with red lines indicating median values and blue caps denoting extrema. Across all models, the median number of techniques per sample lies between 8 and 10, indicating that most generated artifacts exhibit multiple behaviors within a single execution trace. The distributions vary across model families: \textit{DeepSeek-Coder} variants show broader spreads, indicating higher variability, while \textit{Qwen2.5-Coder} and \textit{StarCoder} models exhibit narrower distributions, suggesting more consistent generation patterns.

\begin{figure}
    \centering
    \includegraphics[width=\columnwidth,keepaspectratio]{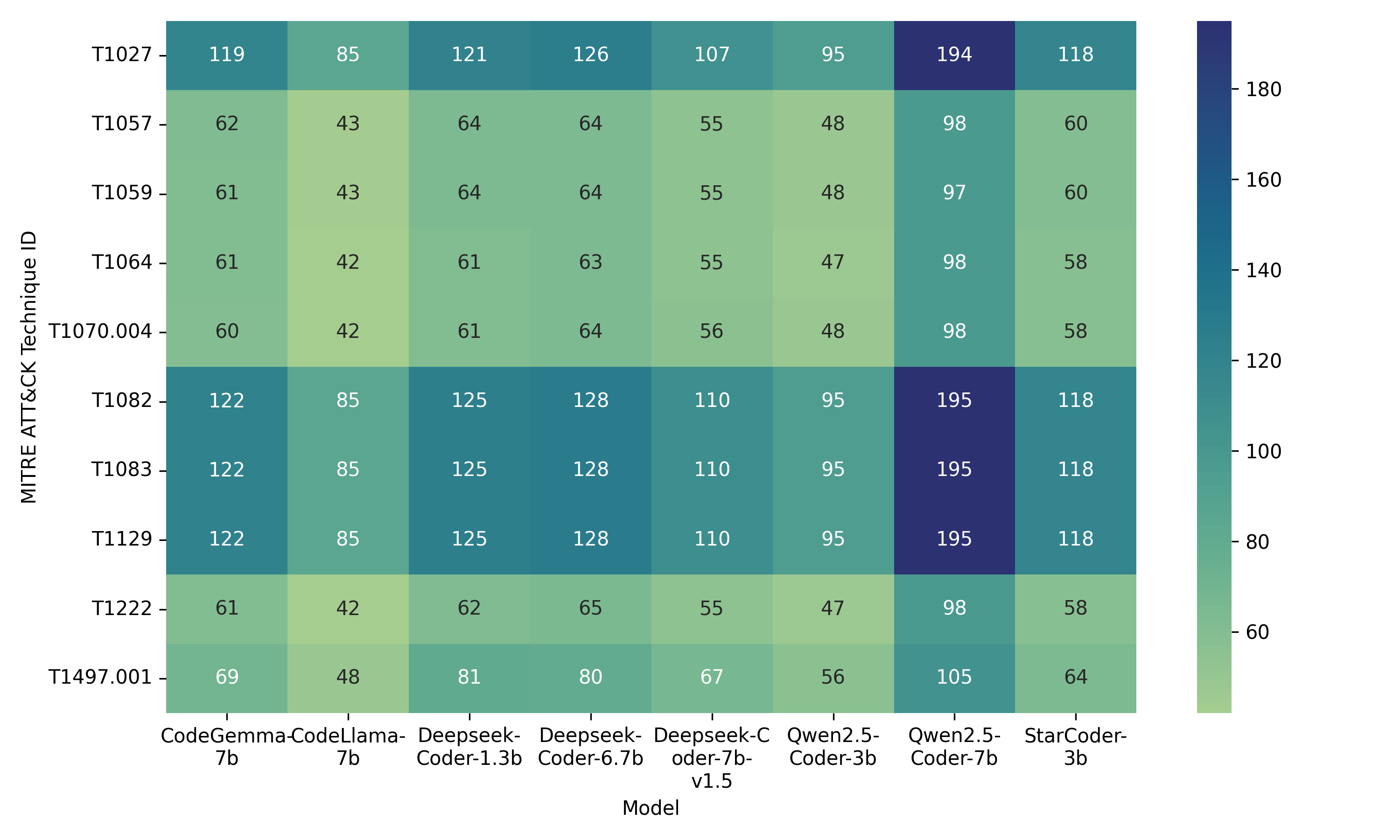}
    \caption{Prevalent techniques across models.}
    \label{fig:top_ttp_heatmap}
\end{figure}

To further examine technique prevalence, Figure~\ref{fig:top_ttp_heatmap} presents the ten most frequently observed ATT\&CK techniques across all models. Techniques such as \texttt{T1027} (Obfuscated Files or Information), \texttt{T1082} (System Information Discovery), and \texttt{T1083} (File and Directory Discovery) appear consistently across multiple models. Differences across model families are also evident: \textit{DeepSeek-Coder} variants and \textit{Qwen2.5-Coder-7B} exhibit broader technique coverage and higher occurrence counts, while \textit{CodeLlama} and \textit{StarCoder} show limited diversity.










Overall, these results indicate that MalGEN generates artifacts with multi-step behavior patterns involving multiple ATT\&CK techniques, while the diversity and frequency of these techniques vary across LLM families.

\begin{center}
\begin{tcolorbox}[
  breakable,
  enhanced,
  colback=gray!5,
  colframe=black,
  title={Key Observation \#2: Technique-level Analysis}
  ]
Generated samples contain multiple ATT\&CK techniques, indicating multi-step behavior patterns. While certain techniques appear consistently across models, the diversity and frequency of techniques vary across LLM families.
\end{tcolorbox}
\end{center}

\subsection{Tactic-Level Analysis}
\label{sec:tactic_analysis_results}

To analyze how generated samples span different stages of the attack lifecycle, we examine both the distribution of distinct MITRE ATT\&CK tactics per sample and the overall tactic coverage across models.

\begin{figure}
    \centering
    \includegraphics[width=\columnwidth]{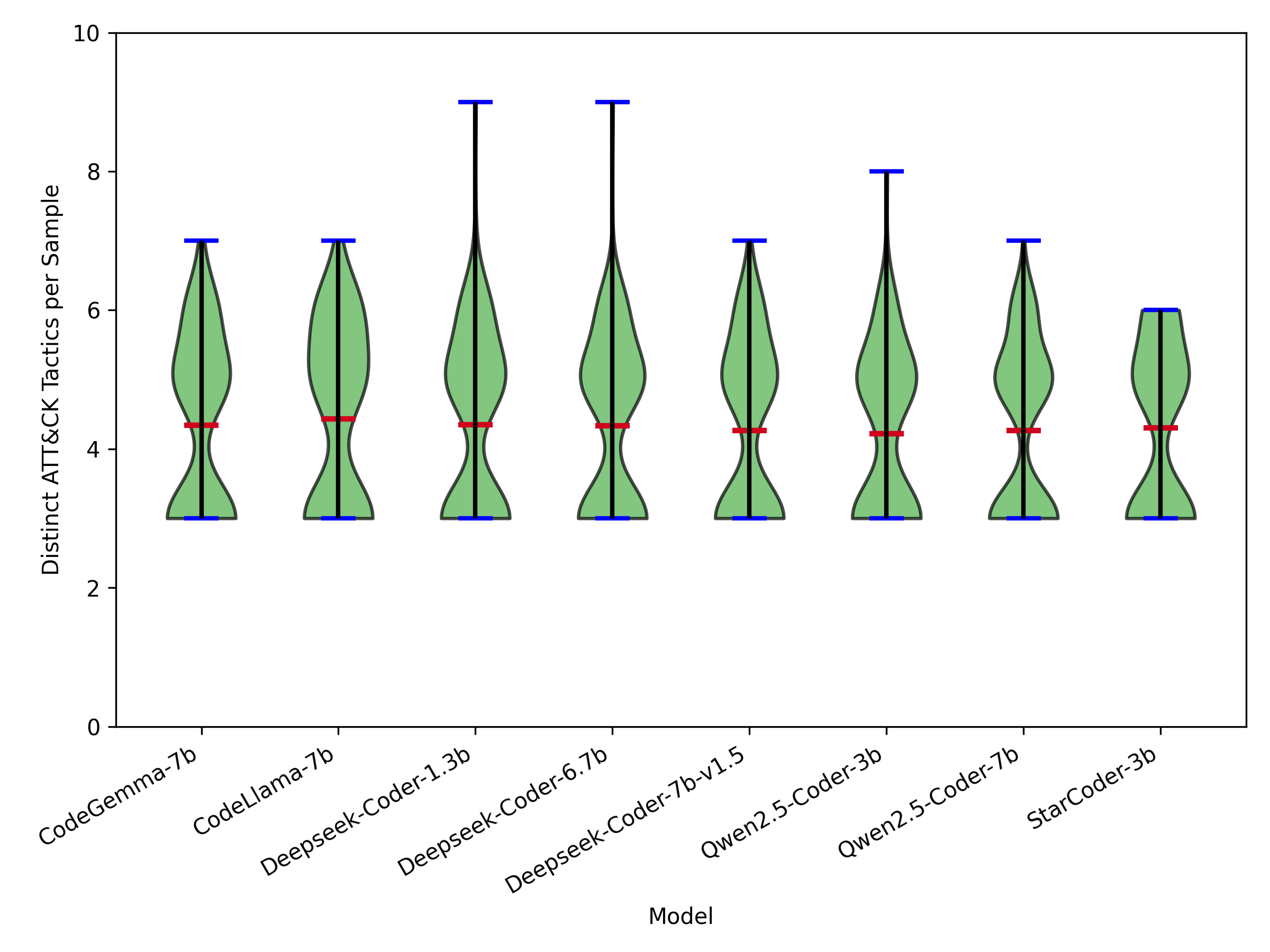}
    \caption{Distribution of tactics per sample.}
    \label{fig:tactic_dist}
\end{figure}

Figure~\ref{fig:tactic_dist} shows the distribution of distinct tactics observed per generated sample across LLM families. Each violin represents the density of tactic counts per sample, with red bars indicating median values and blue caps denoting extrema. The median number of tactics per sample lies between four and five, indicating that most generated artifacts exhibit multi-stage behaviors rather than isolated actions. Variations across model families are evident: \textit{DeepSeek-Coder} variants show broader distributions, indicating higher variability, while models such as \textit{CodeGemma-7B} and \textit{CodeLlama-7B} exhibit narrower distributions, suggesting more consistent patterns.

\begin{figure}
    \centering
    \includegraphics[width=\columnwidth]{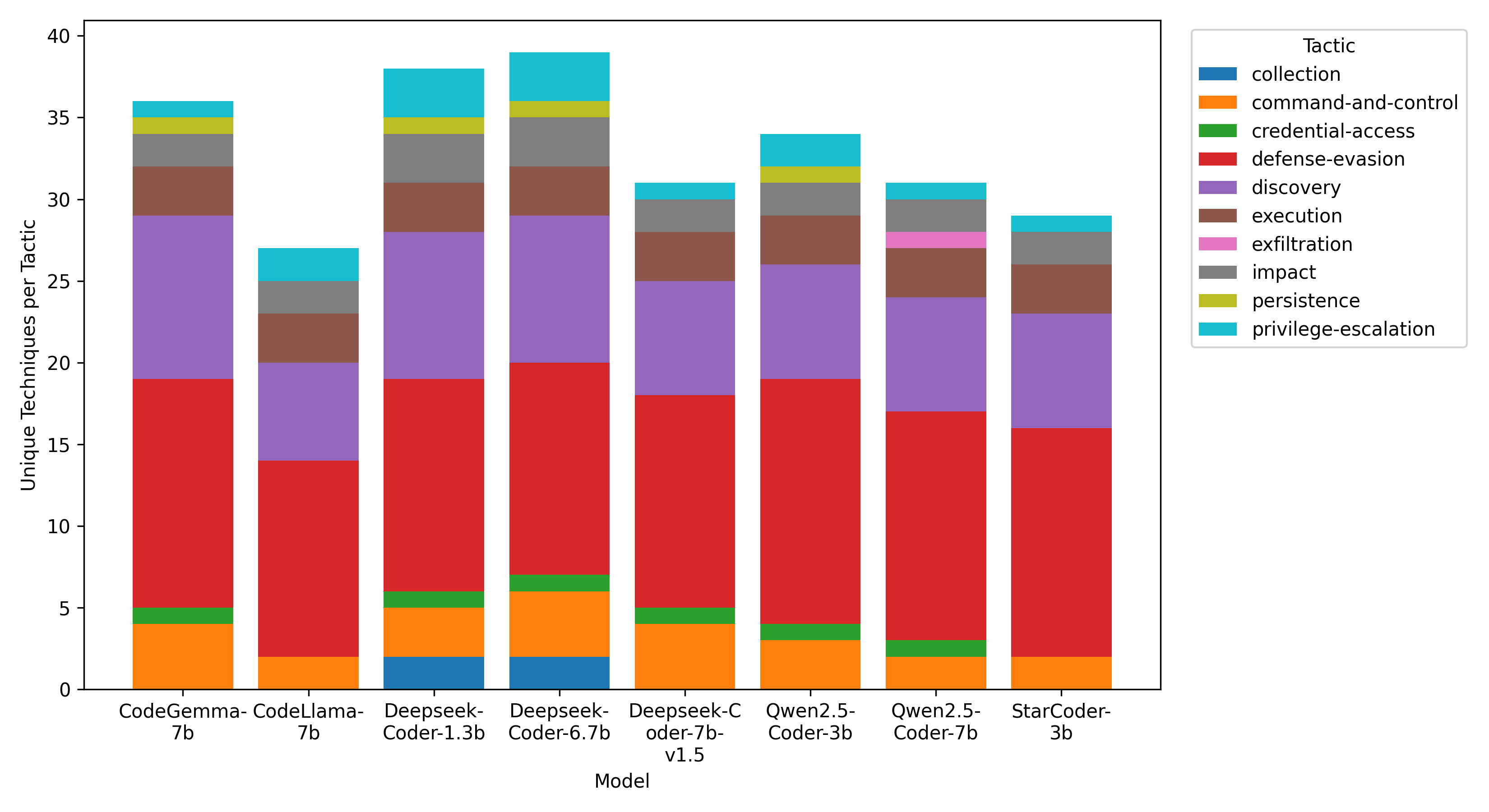}
    \caption{Coverage of tactics across models.}
    \label{fig:tactic_coverage}
\end{figure}

Figure~\ref{fig:tactic_coverage} illustrates tactic coverage across models. \textit{DeepSeek-Coder-6.7B} and \textit{DeepSeek-Coder-1.3B} demonstrate the widest coverage, each expressing close to 40 distinct ATT\&CK techniques, followed by \textit{CodeGemma-7B} and \textit{Qwen2.5-Coder-7B}. In contrast, \textit{CodeLlama-7B} exhibits more limited coverage. Across models, \textit{defense evasion} and \textit{discovery} are most frequently represented, while \textit{execution} and \textit{impact} show moderate presence. Tactics such as \textit{collection}, \textit{exfiltration}, and \textit{command and control} appear less frequently, likely due to the absence of external infrastructure in the experimental setup.

Overall, these results indicate that generated samples commonly exhibit multi-stage behavior patterns spanning multiple ATT\&CK tactics, with both the breadth and composition of tactics varying across LLM families.





\begin{center}
\begin{tcolorbox}[
  breakable,
  enhanced,
  colback=gray!5,
  colframe=black,
  title={Key Observation \#3: Tactic-level Analysis}
  ]
Generated samples commonly exhibit multi-stage behavior patterns spanning multiple ATT\&CK tactics, while the breadth and distribution of tactics vary across LLM families.
\end{tcolorbox}
\end{center}

\subsection{Multi-Stage Prevalence Analysis}
\label{sec:multistage_prevalence_results}

\begin{figure}
    \centering
    \includegraphics[width=\columnwidth]{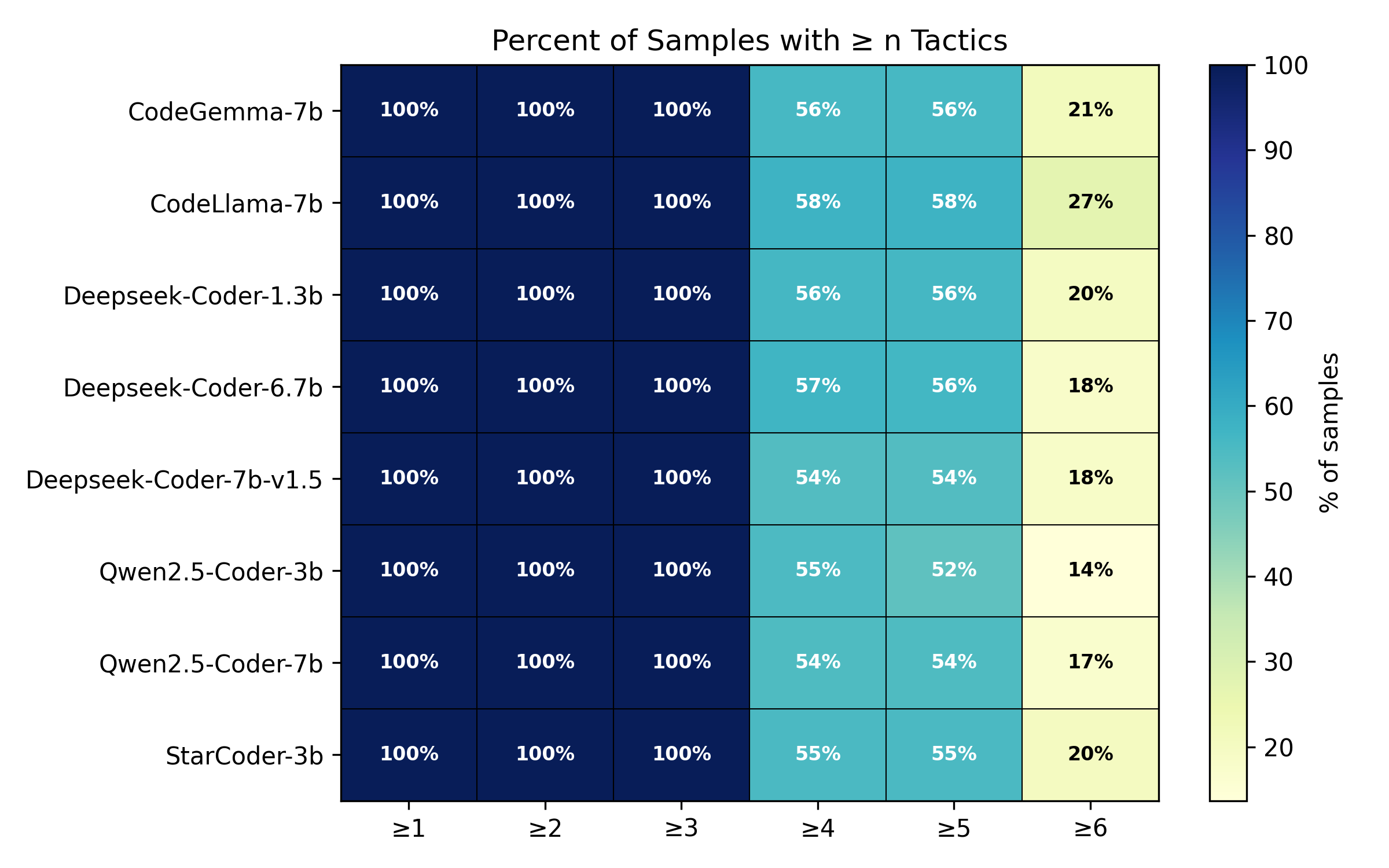}
    \caption{Percentage of generated samples containing at least $n$ distinct ATT\&CK tactics ($n \geq 1$\,–\,6) across LLM variants. 
    }
    \label{fig:n_tactic_heatmap}
\end{figure}

To further examine multi-stage behavior, we analyze the prevalence of samples containing multiple ATT\&CK tactics within a single execution trace. Figure~\ref{fig:n_tactic_heatmap} summarizes this as the percentage of samples containing at least $n$ distinct tactics, where $n \geq 1$\,–\,6, across LLM variants. All generated samples include multiple tactics, with each sample exhibiting at least three distinct tactics. A substantial proportion of samples across models contain four or more tactics, indicating the presence of multi-stage behavior patterns.

Differences across models are also observed. \textit{CodeLlama-7B} shows the highest proportion of samples with six or more tactics (above $25\%$), while \textit{CodeGemma-7B}, \textit{DeepSeek-Coder-1.3B}, and \textit{StarCoder-3B} exhibit similar proportions at around $20\%$. Other models show slightly lower prevalence of such higher-tactic samples.
These results indicate that the generated artifacts frequently combine multiple tactics within a single sample. The extent of this combination varies across models, reflecting differences in how behaviors are composed.

\begin{center}
\begin{tcolorbox}[
  breakable,
  enhanced,
  colback=gray!5,
  colframe=black,
  title={Key Observation \#4: Multi-stage Analysis}
  ]
A substantial portion of generated samples include multiple ATT\&CK tactics, indicating multi-stage behavior patterns with variation across models.
\end{tcolorbox}
\end{center}

\subsection{Detection Coverage Analysis}
\label{sec:detection_coverage_results}

\begin{figure}
    \centering
    \includegraphics[width=\columnwidth]{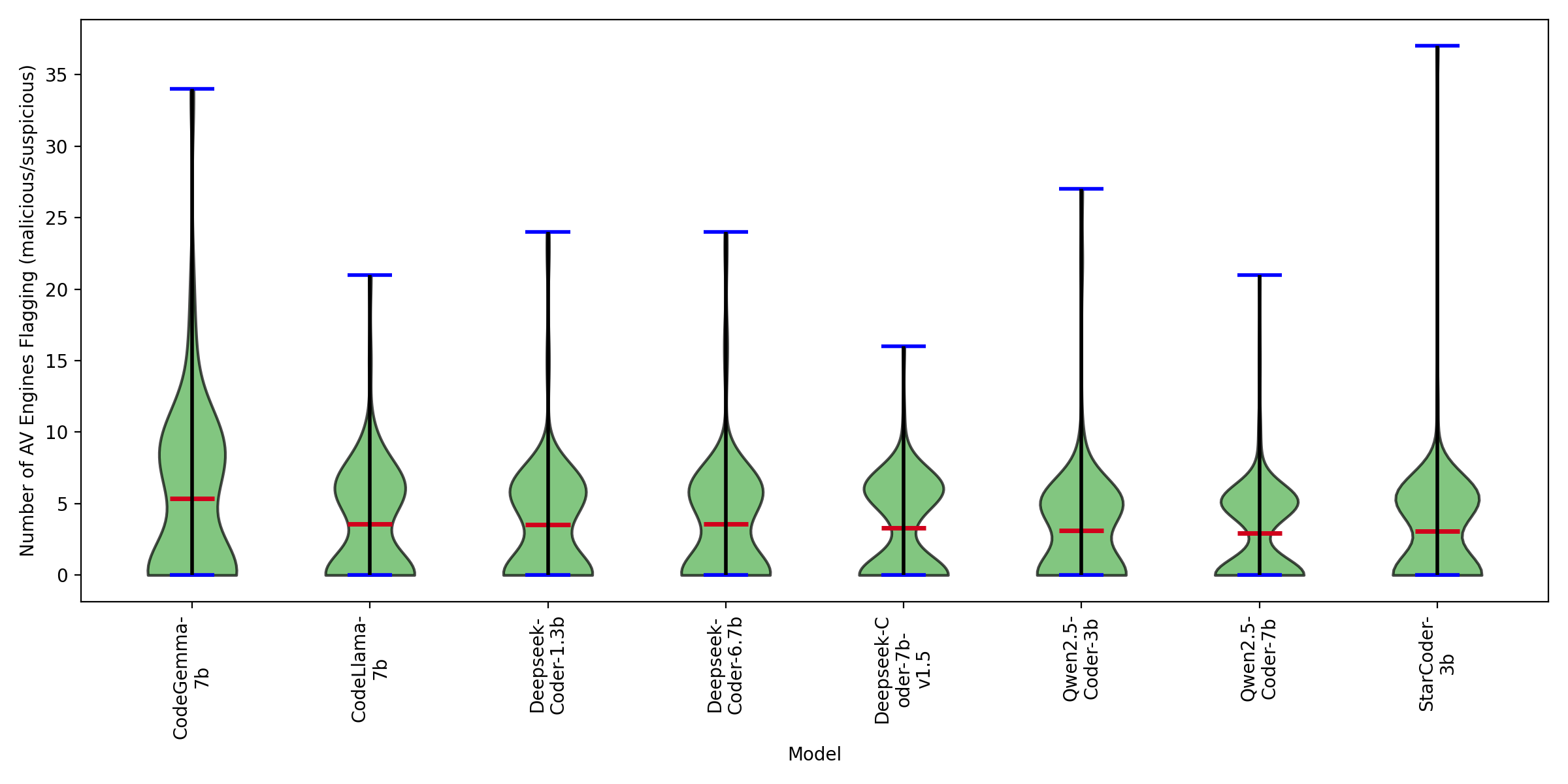}
    \caption{Distribution of detection coverage across VirusTotal engines. 
    Red bars denote medians and blue caps denote extrema.}
    \label{fig:detection_violin}
\end{figure}

To assess how generated artifacts are handled by existing detection systems, all samples were analyzed using VirusTotal (VT), which aggregates results from over $75$ commercial and open-source antivirus engines. Table~\ref{tab:vt_detection_summary} summarizes detection statistics across LLM variants, and Figure~\ref{fig:detection_violin} shows the per-sample distribution of detection counts. It is evident that that most samples receive between zero and five detections, with only a small fraction reaching higher detection counts.

As shown in Table~\ref{tab:vt_detection_summary}, the proportion of samples not flagged by any engine ranges between $43\%$ and $48\%$ across models. Mean detection rates remain low (0.038–0.071), indicating limited agreement across engines even when a sample is detected. Median detection counts fall between 3 and 5 engines per sample, while a small number of samples exhibit higher detection counts, reflected in the long-tailed distributions.

\begin{table*}[t]
\centering
\caption{VirusTotal Detection Summary Across LLM Variants}
\label{tab:vt_detection_summary}
\footnotesize
\begin{tabular}{|c|c|c|c|c|c|}
\hline
\textbf{Model} 
& \makecell{\textbf{\#Samples}} 
& \makecell{\textbf{Undetected (\%)}} 
& \makecell{\textbf{Mean} \textbf{Det. Rate}} 
& \makecell{\textbf{Median} \textbf{Det. Count}} 
& \makecell{\textbf{Max} \textbf{Detections}} \\
\midrule
\hline
CodeGemma-7B         & 122 & 43.44 & 0.0711 & 5.0 & 34 \\

CodeLlama-7B         & 85  & 45.88 & 0.0468 & 3.0 & 21 \\

DeepSeek-Coder-1.3B  & 125 & 44.00 & 0.0462 & 3.0 & 24 \\

DeepSeek-Coder-6.7B  & 127 & 44.09 & 0.0470 & 3.0 & 24 \\

DeepSeek-Coder-7B-v1.5 & 110 & 45.45 & 0.0433 & 3.5 & 16 \\

Qwen2.5-Coder-3B     & 95  & 47.37 & 0.0412 & 4.0 & 27 \\

Qwen2.5-Coder-7B     & 195 & 47.18 & 0.0383 & 3.0 & 21 \\

StarCoder-3B         & 118 & 48.31 & 0.0404 & 3.5 & 37 \\
\midrule
\hline
\textbf{Overall (Average)} & \textbf{977} & \textbf{45.71} & \textbf{0.0468} & \textbf{3.5} & \textbf{25.5} \\
\hline
\end{tabular}
\end{table*}


Differences across models are also observed. For example, \emph{CodeGemma-7B} shows higher average detection counts, while \emph{Qwen2.5-Coder-7B} exhibits comparatively lower detection rates and a higher proportion of samples not flagged by any engine. \emph{StarCoder-3B} shows the largest upper tail, indicating occasional cases with strong multi-engine agreement. The findings indicate that the generated artifacts are not consistently aligned with patterns recognized by existing detection systems.


Importantly, low detection rates do not imply that the generated samples are benign. Our prior analysis of ATT\&CK techniques and tactics shows that these samples exhibit multi-stage malicious behaviors. The low detection rates therefore reflect structural differences in the generated behaviors, which are not adequately captured by existing detection systems.


\begin{center}
\begin{tcolorbox}[
  breakable,
  enhanced,
  colback=gray!5,
  colframe=black,
  title={Key Observation \#5: Detection Coverage Analysis}
  ]
Many generated samples are not consistently flagged by antivirus engines, suggesting that detection outcomes do not fully reflect their underlying behavior.
\end{tcolorbox}
\end{center}

\subsection{Dynamic Behavior-Based Detection Analysis}
\label{subsec:dynamic_detection}



We further extend the analysis to behavior-based detection using traditional machine learning models. Since the samples target both Linux and Windows, we evaluate dynamic detection separately on each platform and report averaged results in Tables~\ref{tab:benchmark_detection} and~\ref{tab:dynamic_detection_results}. On the held-out test set (Table~\ref{tab:benchmark_detection}), all models perform strongly, with accuracy ranging from 0.85 to 0.94, indicating clear separation between benign and malicious behavior.

\begin{table}[h]
\centering
\caption{Detection Performance on Held-out Test Set}
\label{tab:benchmark_detection}
\begin{tabularx}{\columnwidth}{|X|l|l|l|l|}
\hline
\textbf{Model} & \textbf{Accuracy} & \textbf{Precision} & \textbf{Recall} & \textbf{F1-score} \\
\hline
\hline
Logistic Regression & 0.89 & 0.90 & 0.88 & 0.89 \\
Random Forest & 0.93 & 0.94 & 0.92 & 0.93 \\
KNN & 0.87 & 0.88 & 0.86 & 0.87 \\
Linear SVM & 0.91 & 0.92 & 0.90 & 0.91 \\
Naive Bayes & 0.85 & 0.86 & 0.84 & 0.85 \\
Decision Tree & 0.88 & 0.89 & 0.87 & 0.88 \\
XGBoost & 0.94 & 0.95 & 0.93 & 0.94 \\
\hline
\end{tabularx}
\end{table}

The trained models are then applied without retraining to system call traces extracted from the generated samples. Table~\ref{tab:dynamic_detection_results} summarizes the detection performance.
\begin{table}[h]
\centering
\caption{Detection Performance on Generated Samples}
\label{tab:dynamic_detection_results}
\begin{tabularx}{\columnwidth}{|X|l|l|l|l|}
\hline
\textbf{Model} & \textbf{Accuracy} & \textbf{Precision} & \textbf{Recall} & \textbf{F1-score} \\
\hline
\hline
Logistic Regression & 0.71 & 0.73 & 0.69 & 0.71 \\
Random Forest & 0.76 & 0.78 & 0.74 & 0.76 \\
KNN & 0.68 & 0.70 & 0.66 & 0.68 \\
Linear SVM & 0.74 & 0.75 & 0.72 & 0.73 \\
Naive Bayes & 0.65 & 0.67 & 0.63 & 0.65 \\
Decision Tree & 0.70 & 0.72 & 0.68 & 0.70 \\
XGBoost & 0.78 & 0.80 & 0.76 & 0.78 \\
\hline
\end{tabularx}
\end{table}
Across all models, detection performance on generated samples lies between 0.65 and 0.78. Although this is lower than the benchmark results, the classifiers still identify a majority of samples as malicious. This confirms that the generated binaries execute meaningful malicious actions that are visible in their runtime behavior.


The observed gap reflects variation in how these behaviors are expressed. The generated samples follow valid adversarial logic, but differ in system call ordering and behavioral composition compared to patterns seen in standard training datasets. As a result, detection models trained on conventional data show reduced confidence when applied to these samples.
These results show that the generated artifacts are both malicious and behaviorally diverse, and they extend beyond the patterns learned from existing datasets.


\begin{center}
\begin{tcolorbox}[
  breakable,
  enhanced,
  colback=gray!5,
  colframe=black,
  title={Key Observation \#6: Dynamic behavior Analysis}
  ]
Behavior-based detectors identify most generated samples as malicious. Lower detection compared to benchmark data reflects variation in behavior patterns and shows limits of traditional ML-based detection models.
\end{tcolorbox}
\end{center}

\subsection{Malignant Capability Analysis}

To analyze the functional behavior of the generated samples, we use MalXCap~\citep{saha2023malxcap}, a multi-label malware capability extractor. Each generated binary is processed to identify malignant capabilities it exhibits during execution. The predicted capabilities are aggregated to compute the percentage of samples exhibiting each capability.

\begin{figure}
    \centering
    \includegraphics[width=\columnwidth]{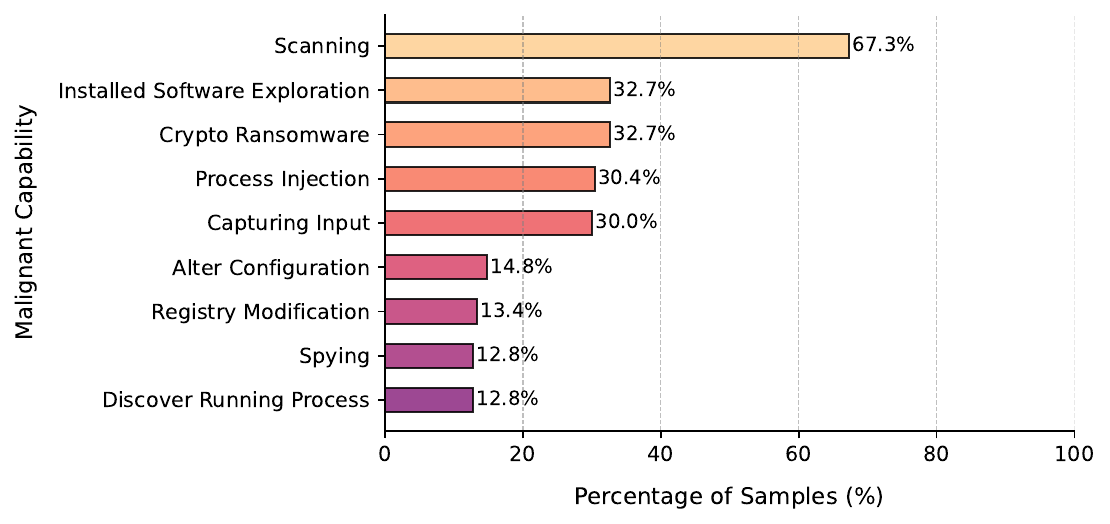}
    \caption{Capability distribution (in \%) detected by MalXCap in generated samples.}
    \label{fig:capibility_analysis}
\end{figure}

As shown in Figure~\ref{fig:capibility_analysis}, Scanning is the most frequently observed capability, appearing in $67.3\%$ of samples. This indicates that many generated artifacts include discovery-related behavior. Other commonly observed capabilities include Installed Software Exploration ($32.7\%$), Crypto-Ransomware ($32.7\%$), and Process Injection ($30.4\%$).
Additional capabilities such as Capturing Input ($30.0\%$), Alter Configuration ($14.8\%$), and Registry Modification ($13.4\%$) are also present, indicating variation in the types of behaviors exhibited across samples. 

These results indicate that the generated artifacts exhibit multiple functional capabilities rather than single isolated behaviors. The distribution across capability categories suggests variability in how different behaviors are combined within individual samples.


\begin{center}
\begin{tcolorbox}[
  breakable,
  enhanced,
  colback=gray!5,
  colframe=black,
  title={Key Observation \#7: Malignant Capability Analysis}
  ]
Generated samples exhibit multiple malignant capabilities, indicating that they combine several behaviors within a single artifact, with variation across capability types.
\end{tcolorbox}
\end{center}

\subsection{Failure Analysis}
\label{subsec:failure_analysis}

Although MalGEN improves executable generation over the single-agent setting, a notable fraction of benchmark settings still fail before producing a valid executable. Out of the 1,920 benchmark settings, 977 result in successful executable generation, while 943 fail during the generation process. These failures arise at different stages of the pipeline, including code synthesis, integration, and packaging. Table~\ref{tab:failure_breakdown} summarizes the observed generation failure categories.

We group generation failures into five categories. \textit{Syntax/build failure} refers to malformed or incomplete code that cannot be compiled or converted into a valid script. \textit{Dependency/import failure} captures missing libraries, unresolved imports, or version mismatches required by the generated code. \textit{Integration conflict} occurs when independently generated modules cannot be combined due to inconsistent interfaces, conflicting identifiers, or broken data flow across sub-tasks. \textit{Packaging failure} corresponds to errors during executable creation, such as PyInstaller build issues or unresolved resource dependencies. \textit{Multi-factor failure} captures cases where logs indicate multiple overlapping issues.

\begin{table}[htbp]
\centering
\caption{Breakdown of generation failure modes across unsuccessful benchmark settings}
\label{tab:failure_breakdown}
\begin{tabular}{lrr}
\toprule
\textbf{Failure Category} & \textbf{\# Cases} & \textbf{Ratio (\%)} \\
\midrule
Syntax/build failure & 248 & 26.30 \\
Dependency/import failure & 196 & 20.79 \\
Integration conflict & 171 & 18.13 \\
Packaging failure & 142 & 15.06 \\
Multi-factor & 186 & 19.72 \\
\midrule
Total & 943 & 100.00 \\
\bottomrule
\end{tabular}
\end{table}

Most failures occur before executable construction. Syntax/build errors, dependency issues, and integration conflicts together account for the majority of unsuccessful cases. This shows that the main challenge lies in producing reliable code across multiple stages rather than in packaging alone. Packaging failures form a smaller but still significant portion, which indicates that deployment into a standalone executable introduces an additional layer of complexity.
The remaining cases fall under unresolved or multi-factor failures, where errors propagate across stages. These cases highlight the complexity of multi-stage generation, where failures are not always attributable to a single component.

Overall, this analysis highlights that the primary limitation of the current pipeline is code reliability across multi-stage generation. It also points to clear areas for improvement, including stronger syntax validation, better dependency handling, and more robust interface alignment during integration.

\section{Limitations and Future Work}
\label{sec:MalGEN_limitations_future_work}

While MalGEN provides a controlled framework for generating and analyzing LLM-driven adversarial artifacts, few limitations remain.
First, the current evaluation is limited to Linux and Windows environments. Although these platforms cover a wide range of use cases, extending the framework to additional systems such as macOS, Android, and embedded or IoT platforms would help assess how the generation pipeline behaves across different execution environments.
Second, the evaluation leverages VirusTotal and sandbox-based analysis to study detection and behavior. Future work could incorporate additional evaluation methods, including controlled deployment scenarios or more detailed behavioral tracing, to better understand detection and execution dynamics.
Third, the current pipeline does not explicitly include mechanisms for automated correction of failed or partially generated samples. Integrating a repair or refinement stage could improve generation reliability and reduce failure cases observed in the experiments.
Overall, these directions highlight opportunities to extend MalGEN in terms of platform coverage, evaluation depth, and generation robustness.

\begin{table*}[t]
\centering
\caption{Representative works on LLM-driven malware generation and evaluation.}
\label{tab:rw_compare}

\small
\setlength{\tabcolsep}{4pt} 

\begin{tabularx}{\textwidth}{p{3.2cm} c p{2.8cm} c c c c c X}
\toprule

\textbf{Study} & \textbf{Year} & \textbf{Generation Type} & \textbf{Auto} & \textbf{Cross} & \textbf{ATT\&CK} & \textbf{AV} & \textbf{\#Samples} & \textbf{Remarks} \\

\midrule

Adamec \& Tur\v{c}an\'{i}k~\citep{adamec2024development} & 2024 &
Prompt-based (HITL) & \xmark & \xmark & \xmark & \pmark &
4 & Proof-of-concept; manual prompting; no behavioral quantification \\

Yamin et al.~\citep{yamin2024combining} & 2024 &
Ransomware Generation & \pmark & \xmark & \xmark & \xmark &
5 & Manual human-in-loop; limited scalability \\

Huang et al.~\citep{huang2024exploring} & 2024 &
Variant Generation & \pmark & \xmark & \xmark & \cmark &
211 & Windows-only; code-level focus \\

He et al.~\citep{he2025large} & 2025 &
Variant Generation & \cmark & \xmark & \xmark & \cmark &
-- & Windows PE-based; static evasion only (no behavioral analysis) \\

Li et al.~\citep{li2025automated} & 2025 &
Variant Generation & \cmark & \xmark & \xmark & \cmark &
$\sim$3.7K/day & Focus on evasion rather than behavior \\

\rowcolor{gray!10}
\textbf{MalGEN (This work)} & \textbf{2025} &
\textbf{Malware Generation} & \textbf{\cmark} & \textbf{\cmark} & \textbf{\cmark} & \textbf{\cmark} &
\textbf{977} & \textbf{End-to-end automated behavior simulation} \\

\bottomrule
\end{tabularx}

\vspace{3pt}
\raggedright\footnotesize
\textit{Legend:} \cmark = supported; \pmark = partial; \xmark = not addressed.

\end{table*}

\section{Related Work}
\label{sec:relatedwork}


Building on prior analyses, this section situates MalGEN within the evolving landscape of LLM-assisted malware generation. Existing research follows two phases: (i) early feasibility studies demonstrating LLM-based malware generation under constrained settings, and (ii) subsequent work on adversarial variants for evaluating detection resilience. We revisit these developments to underscore the gap addressed by MalGEN.

\textbf{Early feasibility studies. }
The initial studies explored whether general-purpose LLMs such as ChatGPT could generate malicious code, payloads, or phishing content under ethical safeguards.
Adamec and Turčaník~\citep{adamec2024development} demonstrated this potential in a hybrid-warfare simulation, where ChatGPT produced phishing emails, macro-embedded documents, and USB-borne ransomware within controlled exercises. Their findings confirmed that LLMs can generate credible attack artifacts, but these experiments were small in scale and lacked behavioral or structural evaluation.
Yamin et al.~\citep{yamin2024combining} further extended this concept by coupling an uncensored model (Dolphin-Mistral) with ChatGPT to co-develop ransomware samples. Their dual-LLM setup achieved functional generation through iterative refinement. Their analysis shows that the feasibility of collaborative LLM roles but still requires human supervision and manual validation.
Collectively, these works demonstrate the feasibility of LLM-generated malware-like code but remain limited in scale, systematic diversity, behavior analysis, and quantitative evaluation.

\textbf{Variant generation and adversarial pipelines.}
The second research wave focused on large-scale and automated variant synthesis, measuring how LLM-generated code affects antivirus (AV) and machine-learning detectors.
Huang et al.~\citep{huang2024exploring} generated 400 malware variants across four categories (worm, ransomware, keylogger, and fileless malware) using GPT-3.5-turbo, of which only 211 could be compiled into functional executables. Their analysis shows that about 38\% of samples evaded AV detection. This marked the first quantitative assessment of LLM-generated malware, showing that minimal LLM-induced code mutations can cause substantial evasion.
Further, He et al.~\citep{he2025large} model LLM-guided malware evasion as a Markov decision process, where the model iteratively applies functionality-preserving modifications to Windows PE binaries to maximize evasion against static detectors.
Li et al.~\citep{li2025automated} extended this paradigm to Android, constructing an adversarial malware factory that merges piggybacking, perturbation generation, and LLM-assisted code synthesis. Their system 
produced thousands of evasive APKs daily, emphasizing how LLMs can accelerate polymorphism and large-scale malware diversification.

\textbf{Research Gaps and Contribution.} While these frameworks demonstrated effective evasion and generation throughput, they primarily focused on producing variants of existing malware samples rather than synthesizing entirely unseen malicious behaviors. Most experiments remained confined to a single platform (either Windows or Android) and did not examine how the generated samples varied in their underlying adversarial techniques. Furthermore, none incorporated structured agentic coordination in the generation process.

MalGEN advances beyond these prior works by introducing a multi-agent, behavior-aware malware generation framework.
It decomposes high-level malicious intents into coordinated sub-tasks, planning, code generation, integration, and evaluation, executed by autonomous LLM agents. Unlike existing pipelines that evaluate code similarity or AV evasion alone, MalGEN quantifies behavioral diversity using MITRE ATT\&CK tactics and techniques. This behavior-grounded approach establishes MalGEN as a testbed for studying LLM-driven malware generation and adversarial behavior at scale.
Table~\ref{tab:rw_compare} provides a comparative overview of MalGEN against notable studies in existing literature.

\section{Conclusion}
\label{sec:conclusion}

MalGEN introduces a novel, ethically grounded testbed for modeling and analyzing AI-generated malware behaviors through coordinated, multi-agent reasoning. By reframing malware generation as a structured adversarial workflow, it enables the creation of diverse, multi-tactic, and behaviorally realistic samples aligned with MITRE ATT\&CK TTPs. Evaluations across eight code-oriented LLM variants demonstrate both the adaptive potential and evasive nature of such artifacts, revealing critical detection blind spots in existing defensive systems. Serving as a reproducible, modular, and safe environment for proactive experimentation, MalGEN bridges the gap between offensive innovation and defensive preparedness, inviting the cybersecurity community to rethink how we train, test, and evolve our defenses in the age of generative intelligence.












\section*{Declaration of Generative AI and AI-assisted technologies in the writing process}
During the preparation of this manuscript, the authors used AI-based tools, including ChatGPT, Gemini, and Claude, to assist with text refinement, sentence rewriting, and grammar and spelling corrections. All outputs generated by these tools were carefully reviewed and edited by the authors. The authors take full responsibility for the content and integrity of the published work.

\printcredits

\bibliographystyle{cas-model2-names}

\bibliography{cas-refs}

@inproceedings{pa2023attacker,
  title={An attacker’s dream? exploring the capabilities of chatgpt for developing malware},
  author={Pa Pa, Yin Minn and Tanizaki, Shunsuke and Kou, Tetsui and Van Eeten, Michel and Yoshioka, Katsunari and Matsumoto, Tsutomu},
  booktitle={Proceedings of the 16th cyber security experimentation and test workshop},
  pages={10--18},
  year={2023}
}

@inproceedings{he2025large,
  title={From Large Language Models to Adversarial Malware: How far are we},
  author={He, Shuai and Yan, Hao and Li, Wenke and Hong, Sheng and Guo, Xiaowei and Liu, Xiaofan and Fu, Cai},
  booktitle={Proceedings of the 34th ACM SIGSOFT International Symposium on Software Testing and Analysis},
  pages={178--182},
  year={2025}
}

@inproceedings{huang2024exploring,
  title={Exploring the Impact of LLM Assisted Malware Variants on Anti-Virus Detection},
  author={Huang, Zhewei and Pa, Yin Minn Pa and Yoshioka, Katsunari},
  booktitle={2024 IEEE Conference on Dependable and Secure Computing (DSC)},
  pages={76--77},
  year={2024},
  organization={IEEE}
}

@techreport{enisa2025threatlandscape,
  title        = {ENISA Threat Landscape 2025},
  author       = {{European Union Agency for Cybersecurity (ENISA)}},
  year         = {2025},
  institution  = {ENISA},
  month        = {October},
  note         = {Available online: \url{https://www.enisa.europa.eu/publications/enisa-threat-landscape-2025}},
}

@inproceedings{hidayat2025advancements,
  title={Advancements in Machine Learning and Deep Learning for Malware Detection Challenges Breakthroughs},
  author={Hidayat, Syamsu and Utami, Ema and Setyanto, Arief and Karim, Abdul and others},
  booktitle={2025 4th International Conference on Creative Communication and Innovative Technology (ICCIT)},
  pages={1--6},
  year={2025},
  organization={IEEE}
}

@article{gorment2023machine,
  title={Machine learning algorithm for malware detection: Taxonomy, current challenges, and future directions},
  author={Gorment, Nor Zakiah and Selamat, Ali and Cheng, Lim Kok and Krejcar, Ondrej},
  journal={IEEE Access},
  volume={11},
  pages={141045--141089},
  year={2023},
  publisher={IEEE}
}

@article{naveed2025comprehensive,
  title={A comprehensive overview of large language models},
  author={Naveed, Humza and Khan, Asad Ullah and Qiu, Shi and Saqib, Muhammad and Anwar, Saeed and Usman, Muhammad and Akhtar, Naveed and Barnes, Nick and Mian, Ajmal},
  journal={ACM Transactions on Intelligent Systems and Technology},
  volume={16},
  number={5},
  pages={1--72},
  year={2025},
  publisher={ACM New York, NY}
}

@article{xu2024large,
  title={Large language models for cyber security: A systematic literature review},
  author={Xu, HanXiang and Wang, ShenAo and Li, Ningke and Wang, Kailong and Zhao, Yanjie and Chen, Kai and Yu, Ting and Liu, Yang and Wang, HaoYu},
  journal={ACM Transactions on Software Engineering and Methodology},
  year={2024},
  publisher={ACM New York, NY}
}

@inproceedings{saha2025parag,
  title={PARAG: P roactive A nswering Framework Integrating LLMs with R etrieval-A ugmented G eneration},
  author={Saha, Bikash and Rani, Nanda and Chakraborty, Joheen and Singh, Divyanshu and Chakraborty, Soumyo V and Shukla, Sandeep Kumar},
  booktitle={European Interdisciplinary Cybersecurity Conference},
  pages={20--37},
  year={2025},
  organization={Springer}
}

@inproceedings{saha2025malaware,
  title={Malaware: Automating the comprehension of malicious software behaviours using large language models (llms)},
  author={Saha, Bikash and Rani, Nanda and Shukla, Sandeep Kumar},
  booktitle={2025 IEEE/ACM 22nd International Conference on Mining Software Repositories (MSR)},
  pages={169--173},
  year={2025},
  organization={IEEE}
}

@article{sheng2025llms,
  title={Llms in software security: A survey of vulnerability detection techniques and insights},
  author={Sheng, Ze and Chen, Zhicheng and Gu, Shuning and Huang, Heqing and Gu, Guofei and Huang, Jeff},
  journal={ACM Computing Surveys},
  year={2025},
  publisher={ACM New York, NY}
}

@article{tellache2025advancing,
  title={Advancing autonomous incident response: Leveraging llms and cyber threat intelligence},
  author={Tellache, Amine and Korba, Abdelaziz Amara and Mokhtari, Amdjed and Moldovan, Horea and Ghamri-Doudane, Yacine},
  journal={arXiv preprint arXiv:2508.10677},
  year={2025}
}

@article{achiam2023gpt,
  title={Gpt-4 technical report},
  author={Achiam, Josh and Adler, Steven and Agarwal, Sandhini and Ahmad, Lama and Akkaya, Ilge and Aleman, Florencia Leoni and Almeida, Diogo and Altenschmidt, Janko and Altman, Sam and Anadkat, Shyamal and others},
  journal={arXiv preprint arXiv:2303.08774},
  year={2023}
}

@article{tayyab2022survey,
  title={A survey of the recent trends in deep learning based malware detection},
  author={Tayyab, Umm-e-Hani and Khan, Faiza Babar and Durad, Muhammad Hanif and Khan, Asifullah and Lee, Yeon Soo},
  journal={Journal of Cybersecurity and Privacy},
  volume={2},
  number={4},
  pages={800--829},
  year={2022},
  publisher={MDPI}
}

@article{aboaoja2022malware,
  title={Malware detection issues, challenges, and future directions: A survey},
  author={Aboaoja, Faitouri A and Zainal, Anazida and Ghaleb, Fuad A and Al-Rimy, Bander Ali Saleh and Eisa, Taiseer Abdalla Elfadil and Elnour, Asma Abbas Hassan},
  journal={Applied Sciences},
  volume={12},
  number={17},
  pages={8482},
  year={2022},
  publisher={MDPI}
}

@article{li2025automated,
  title={Automated Mass Malware Factory: The Convergence of Piggybacking and Adversarial Example in Android Malicious Software Generation},
  author={Li, Heng and Yao, Zhiyuan and Wu, Bang and Gao, Cuiying and Xu, Teng and Yuan, Wei and Luo, Xiapu}
}

@inproceedings{yamin2024combining,
  title={Combining Uncensored and Censored LLMs for Ransomware Generation},
  author={Yamin, Muhammad Mudassar and Hashmi, Ehtesham and Katt, Basel},
  booktitle={International Conference on Web Information Systems Engineering},
  pages={189--202},
  year={2024},
  organization={Springer}
}

@inproceedings{adamec2024development,
  title={Development of Malware Using Large Language Models},
  author={Adamec, Matej and Tur{\v{c}}an{\'\i}k, Michal},
  booktitle={2024 New Trends in Signal Processing (NTSP)},
  pages={1--5},
  year={2024},
  organization={IEEE}
}

@inproceedings{madani2023metamorphic,
  title={Metamorphic malware evolution: The potential and peril of large language models},
  author={Madani, Pooria},
  booktitle={2023 5th IEEE International Conference on Trust, Privacy and Security in Intelligent Systems and Applications (TPS-ISA)},
  pages={74--81},
  year={2023},
  organization={IEEE}
}

@article{beckerich2023ratgpt,
  title={Ratgpt: Turning online llms into proxies for malware attacks},
  author={Beckerich, Mika and Plein, Laura and Coronado, Sergio},
  journal={arXiv preprint arXiv:2308.09183},
  year={2023}
}

@incollection{strom2018mitre,
  title={Mitre att\&ck: Design and philosophy},
  author={Strom, Blake E and Applebaum, Andy and Miller, Doug P and Nickels, Kathryn C and Pennington, Adam G and Thomas, Cody B},
  booktitle={Technical report},
  year={2018},
  publisher={The MITRE Corporation}
}

@article{shevlane2023model,
  title={Model evaluation for extreme risks},
  author={Shevlane, Toby and Farquhar, Sebastian and Garfinkel, Ben and Phuong, Mary and Whittlestone, Jess and Leung, Jade and Kokotajlo, Daniel and Marchal, Nahema and Anderljung, Markus and Kolt, Noam and others},
  journal={arXiv preprint arXiv:2305.15324},
  year={2023}
}

@article{brundage2018malicious,
  title={The malicious use of artificial intelligence: Forecasting, prevention, and mitigation},
  author={Brundage, Miles and Avin, Shahar and Clark, Jack and Toner, Helen and Eckersley, Peter and Garfinkel, Ben and Dafoe, Allan and Scharre, Paul and Zeitzoff, Thomas and Filar, Bobby and others},
  journal={arXiv preprint arXiv:1802.07228},
  year={2018}
}

@inproceedings{saha2023malxcap,
  title={MalXCap: A Method for Malware Capability Extraction},
  author={Saha, Bikash and Rani, Nanda and Shukla, Sandeep Kumar},
  booktitle={International Conference on Information Security Practice and Experience},
  pages={230--249},
  year={2023},
  organization={Springer}
}

@data{tqqm-aq14-19,
doi = {10.21227/tqqm-aq14},
url = {https://dx.doi.org/10.21227/tqqm-aq14},
author = {Angelo Oliveira},
publisher = {IEEE Dataport},
title = {Malware Analysis Datasets: API Call Sequences},
year = {2019} }

@data{ygtk-hv95-25,
doi = {10.21227/ygtk-hv95},
url = {https://dx.doi.org/10.21227/ygtk-hv95},
author = {Jaubert Long},
publisher = {IEEE Dataport},
title = {Cross-Architecture ELF Malware Dataset (ARM/x86/x64)},
year = {2025} }

@misc{pikker_sandbox,
  author       = {{Cuckoo Sandbox}},
  title        = {Pikker Malware Analysis Sandbox},
  howpublished = {\url{https://sandbox.pikker.ee}},
  note         = {Online malware analysis sandbox based on Cuckoo Sandbox},
  year         = {n.d.},
}

@article{geng2024survey,
  title={A survey of strategy-driven evasion methods for PE malware: Transformation, concealment, and attack},
  author={Geng, Jiaxuan and Wang, Junfeng and Fang, Zhiyang and Zhou, Yingjie and Wu, Di and Ge, Wenhan},
  journal={Computers \& Security},
  volume={137},
  pages={103595},
  year={2024},
  publisher={Elsevier}
}

@article{afianian2019malware,
  title={Malware dynamic analysis evasion techniques: A survey},
  author={Afianian, Amir and Niksefat, Salman and Sadeghiyan, Babak and Baptiste, David},
  journal={ACM Computing Surveys (CSUR)},
  volume={52},
  number={6},
  pages={1--28},
  year={2019},
  publisher={ACM New York, NY, USA}
}

@inproceedings{jordaney2017transcend,
  title={Transcend: Detecting concept drift in malware classification models},
  author={Jordaney, Roberto and Sharad, Kumar and Dash, Santanu K and Wang, Zhi and Papini, Davide and Nouretdinov, Ilia and Cavallaro, Lorenzo},
  booktitle={26th USENIX security symposium (USENIX security 17)},
  pages={625--642},
  year={2017}
}

@inproceedings{barbero2022transcending,
  title={Transcending transcend: Revisiting malware classification in the presence of concept drift},
  author={Barbero, Federico and Pendlebury, Feargus and Pierazzi, Fabio and Cavallaro, Lorenzo},
  booktitle={2022 IEEE Symposium on Security and Privacy (SP)},
  pages={805--823},
  year={2022},
  organization={IEEE}
}

@article{demetrio2021adversarial,
  title={Adversarial exemples: A survey and experimental evaluation of practical attacks on machine learning for windows malware detection},
  author={Demetrio, Luca and Coull, Scott E and Biggio, Battista and Lagorio, Giovanni and Armando, Alessandro and Roli, Fabio},
  journal={ACM Transactions on Privacy and Security (TOPS)},
  volume={24},
  number={4},
  pages={1--31},
  year={2021},
  publisher={ACM New York, NY, USA}
}



\end{document}